\documentclass[amsmath,amssymb,aps,twocolumn,nofootinbib,showpacs,pra,floatfix]{revtex4-1}
\usepackage{graphics} 
\usepackage{amsmath}
\usepackage{psfrag}
\usepackage{graphicx}
\usepackage{hyperref}
\newcommand{\ve}[1]{\mathbf{q1}}

\def\de{\delta}

\def\vep{\varepsilon}

\def\ga{\gamma}

\def\be{\begin{equation}} 
\def\ee{\end{equation}} 
\def\bea{\begin{eqnarray}} 
\def\eea{\end{eqnarray}}  
\def\bean{\begin{eqnarray*}} 
\def\eean{\end{eqnarray*}} 
\def\dd{\partial}

\def\bse{\begin{subequations}}
\def\ese{\end{subequations}}

\def\lsim{\raise 0.4ex\hbox{$<$}\kern -0.8em\lower 0.62ex\hbox{$\sim$}} 
\def\gsim{\raise 0.4ex\hbox{$>$}\kern -0.7em\lower 0.62ex\hbox{$\sim$}}

\def\f0N{f_0^{(N)}}
\def\bec{\begin{center}}
\def\eec{\end{center}}

\def\bg{{\mathbf g}}

\def\bx{{\bf x}}  
\def\bv{{\bf v}}

\def\by{{\bf y}}
\def\bz{{\bf z}}

\def\bF{{\mathbf F}}

\def\mB{\mathcal{B}}
\def\mC{\mathcal{C}}

\def\mS{\mathcal{S}}

\begin{document} 

\title{Finite-$N$ corrections to Vlasov dynamics and the range of pair interactions}
\author{Andrea Gabrielli$^{1,2}$, Michael Joyce$^{3,4}$ and Jules Morand$^{3,4}$} 
\affiliation{$^1$ ISC - CNR, UoS ``Sapienza'', Dipartimento di Fisica, Universit\`a ``Sapienza", Piazzale Aldo Moro 5, 00185 - Rome,Italy}
\affiliation{$^2$INFM, Unit\`a Roma 1, Dipartimento di Fisica, Universit\`a ``Sapienza", Piazzale Aldo Moro 5, 00185 - Rome,Italy}
\affiliation{$^{3}$UPMC Univ Paris 06, UMR 7585, LPNHE, F-75005, Paris, France}
\affiliation{$^{4}$CNRS IN2P3, UMR 7585, LPNHE, F-75005, Paris, France}

\begin{abstract}   
\begin{center}    
{\large\bf Abstract}
\end{center}   
We explore the conditions on a pair interaction for the validity of the 
Vlasov equation to describe the dynamics of an interacting $N$ particle 
system in the large $N$ limit.  Using a coarse-graining in phase
space of the exact Klimontovich equation for the $N$ particle system,
we evaluate, neglecting correlations of density fluctuations,  
the scalings with $N$ of the terms describing the corrections to the 
Vlasov equation for the coarse-grained one particle phase 
space density. Considering a generic 
interaction with radial pair force $F(r)$, with  $F(r) \sim 1/r^\gamma$ at large scales, and 
regulated to a bounded behaviour below a ``softening" 
scale $\varepsilon$, we find that there is an essential 
qualitative difference between the cases $\gamma < d$ 
and $\gamma > d$, i.e., depending on the integrability 
at large distances of the pair force. In the former case the 
corrections to the Vlasov dynamics for a given coarse-grained 
scale are essentially insensitive to the softening  parameter $\varepsilon$,  
while for $\gamma > d$ the amplitude of these terms is 
directly regulated by $\varepsilon$, and thus by the small
scale properties of the interaction.
This corresponds to a simple physical criterion for 
a basic distinction between long-range ($\gamma \leq d $) and 
short range  ($\gamma > d$) interactions, different to the 
canonical one ($\gamma \leq d +1$ or $\gamma > d +1$ ) based
on thermodynamic analysis. This alternative classification, based 
on purely dynamical considerations, is relevant notably to
understanding the conditions for the existence of so-called 
quasi-stationary states in long-range interacting systems. 
\end{abstract}

\maketitle   
\date{today}  

\twocolumngrid   

Interactions are canonically characterized as short-range or long-range   
on the basis of the fundamental distinction which arises in equilibrium 
statistical mechanics between interactions for which the energy is additive 
and those for which it is non-additive (for reviews, see e.g. 
\citep{Campa2009,Dauxois2002,Campa2008,Bouchet2010}).
For a system of particles interacting
via two body interactions with a pair potential $V(r)$, the system  is
then long-range (or ``strongly long range''  \citep{Bouchet2010} )
if and only if $V(r)$ decays at large distances slower than one over the separation $r$ to the power of the 
spatial dimension $d$. In the last decade there has been considerable study of this class of
interactions. One of the very interesting results about systems in
this class which has emerged --- essentially through numerical study of different
models  --- is that, like for the much studied case of gravity
in astrophysics, their dynamics leads, from generic initial conditions,
to so-called quasi-stationary states: macroscopic non-equilibrium states 
which evolve only on timescales which diverge with particle number.
Theoretically these states are interpreted in terms of a description
of the system's dynamics by the Vlasov equation, of which they
represent stationary solutions.  Their physical realizations 
arise in numerous and very diverse systems, ranging from
galaxies and ``dark matter halos" in astrophysics and cosmology 
(see e.g. \cite{Binney2008})
to the red spot on Jupiter (see e.g. \cite{Bouchet2002}), to 
laboratory systems such as cold atoms \cite{Olivetti2009}, and even to
biological systems\cite{Sopik2005}. A basic question is whether the 
appearance of these out of equilibrium stationary
states --- and more generally the validity of the Vlasov equation 
to describe the system's dynamics 
---  applies to the same class of long-range interactions as defined by
equilibrium statistical mechanics, or only to a sub-class of them,   
or indeed to a larger class of interactions. In 
short, in what class of systems can we expect to see
these quasi-stationary states? Are they typical of 
long-range interactions as defined canonically? Or are 
they characteristic of a different class? 

To answer these questions requires establishing the conditions 
of validity of the Vlasov equation, and specifically how such
conditions depend on the two-body interaction. In  the literature
there are, on the one hand, some rigourous mathematical 
results establishing sufficient conditions for the existence
of the Vlasov limit. It has been proven notably 
\citep{Braun1977,Spohn1991,Hauray2007} that the Vlasov equation is valid 
on times scales of order $\sim log N$ times the dynamical time, for 
strictly bound pair potentials decaying at large separations $r$
slower than $r^{-(d-2)}$. On the other hand, both 
results of numerical study and various theoretical approaches, 
based on different approximations or assumptions, 
suggest that much weaker conditions are sufficient,
and the timescales for the validity of the Vlasov
equation can be much longer. 
In the much studied case of gravity, notably, a 
treatment originally introduced by Chandrasekhar,
\citep{Chandrasekhar1943}
and subsequently refined by other authors (see e.g.
\citep{Henon1958,Chavanis2010,Farouki1994,Farouki1982})
in which non-Vlasov effects are assumed to be
dominated by incoherent two body interactions
gives a time scale $\sim N/log N$ times
the dynamical time for the validity
of the Vlasov equation, at least close to 
stationary solutions representing quasi-stationary
states, and this in absence of
a regularisation of the singularity in the two-body 
potential. Theoretical approaches in the
physics literature derive the Vlasov equation
and kinetic equations describing corrections to
it (for a review see\citep{Campa2009,Chavanis2012a}
either within the framework of the BBGKY 
hierarchy\citep{balescu1975}  or starting from the 
exact Klimontovich equation for the N
body system \citep{Klimontovich1967,Nicholson1983}). 
These approaches are both widely argued (see e.g. 
\citep{Chavanis2012a,Chavanis2013,Campa2009,Bouchet2010,Filho2012}), 
to lead generally to lifetimes of quasi-stationary 
states of order $\sim N$ times the dynamical time
for any softened pair potential , except in the
special case of spatially homogeneous quasi-stationary 
states one dimension.  

In this article we address the question of 
the validity of Vlasov dynamics using an approach 
starting from the exact Klimontovich equation.
Instead of considering, as is often done
(see e.g. \cite{Campa2009, Chavanis2012a,Bouchet2010}), an average over 
an ensemble of initial conditions to define a smooth 
one particle phase density, we follow an approach
(described e.g. in \cite{Buchert2005}) in which
such a smoothed density is obtained by 
performing  a coarse-graining in 
phase space.  This approach gives the Vlasov equation for the 
coarse-grained phase space density when certain terms are 
discarded.  We then study how the latter ``non-Vlasov" terms depend
on the particle number $N$, and on the scales introduced
by the coarse-graining. In particular we develop this study 
analyzing the dependence on the large and small distance 
behavior of the two body potential.
Our analysis leading to the scaling behaviours
of these terms is based only one very simple --- 
but physically reasonable --- hypothesis that
we can neglect all correlations in the (microscopic) 
$N$ body configurations other than those coming 
from the mean (coarse-grained) phase space 
density.
The main physical result we highlight is that,
under this simple hypothesis, the coarse-grained 
dynamics of an interacting $N$-particle system 
shows a very different dependence on the pair interaction 
at small scale depending on how fast the interaction decays at 
large distances: for interactions of which the 
{\it pair force is integrable at large scales} the 
coarse-grained $N$ body dynamics is
highly sensitive to how the potential
is softened at much smaller scales,
while for pair forces which are 
non-integrable the opposite is true.
Correspondingly, while the Vlasov limit may be 
obtained for {\it any} pair interaction which is
softened suitably at small scales, the conditions
on the short-scale behaviour of the interaction
are very different depending on whether
its large scale behaviour is in one of
of these two classes. This result provides 
a more rigorous basis for a ``dynamical 
classification" of interactions as long-range
or short-range, which has been introduced
on the basis of simple considerations of
the probability distribution of the force on
a random particle in a uniform particle
distribution in \citep{Gabrielli2010}, 
and found also in \citep{Gabrielli2010a} to coincide 
with a classification based on the dependence on 
softening of collisional time scales using a generalisation
of the analysis of Chandrasekhar for the case
of gravity.

The article is organized as follows. In the next section we 
derive the equation for the coarse-grained phase space
density and write in a simple form the non-Vlasov terms 
which our subsequent analysis focusses on. In section 
\ref{Statistical evaluation of the finite-$N$ fluctuating terms}
we first explain our central hypothesis concerning the
$N$-body dynamics, and then apply it to evaluate the
statistical properties of the non-Vlasov terms.
In the following section we then determine the
scaling behaviours of these expressions, i.e., how
they depend parametrically on the relevant 
parameters introduced, and in particular on
the two parameters characterising the two body 
interaction --- its large scale decay and the scale
at which it is softened.  In section 
\ref{Force fluctuations about the Vlasov limit}
we use these expressions to identify the 
dominant contributions to the non-Vlasov
terms, which turn out to differ depending on how rapidly
the interaction decays at large scales. In 
the following section we present more complete
exact results for the one dimensional case and 
the comparison with a simple numerical 
simulation. We then summarize our results
and conclusions, discussing in particular the
central assumptions and the dependence of
our findings on them.

\section{A Vlasov-like equation for the coarse-grained phase space density}

In this section we summarize an approach used 
to justify the validity of the Vlasov equation for long-range interacting systems.
The approach involves using a coarse-graining, in phase space, of the full N 
body dynamics and leads to an evolution equation for the coarse grained
phase space density which consists of the Vlasov terms, plus additional
terms. This equation, and the specific form of the non-Vlasov terms
we derive, is the starting point for our analysis in the subsequent 
sections. We follow closely at the beginning the presentation 
and notation of  \cite{Buchert2005}. 

We consider a $d$-dimensional system of $N$ particles of identical mass $m=1$ 
interacting only through the a generic two body force, denoting $\bg (\bx)$ the 
force on a particle at $\bx$ exerted by another one at the origin.

At any time $t$, the $N$ particles have phase space positions which
we denote $\{(\bx_{i},\bv_{i})\}_{i=1..N}$, and the  microscopic 
(or fine-grained, or Klimontovich) one particle phase-space 
density is simply the distribution 
\be
f_{k}(\bx,\bv,t)=\sum_{i=1}^{N}\de(\bx-\bx_{i}(t))\de(\bv-\bv_{i}(t))\,.
\ee
Likewise the microscopic density in one particle coordinate space is
\be 
n_{k}(\bx)=\int f_{k}d^{d}\bv=\sum_{i=1}^{N}\de(\bx-\bx_{i})
\ee

The full evolution of the $N$ body system can be written in
the form of the so-called Klimontovich equation for the 
microscopic phase space density: 
\be \label{eq:Klimontovich}
\frac{\dd f_{k}}{\dd t } + \bv \frac{\dd f_{k}}{\dd \bx} + \bF[n_k](\bx) \frac{\dd f_{k}}{\dd \bv} = 0.
\ee 
where
\be
\bF[n_k](\bx)=\int_{\Omega}\bg(\bx-\bx')n_{k}(\bx')d^{d}\bx'=\sum_{i=1}^{N}\bg(\bx-\bx_{i})
\ee
is the exact force at point $x$ (due to all other particles). The only assumption made in
deriving this equation from the equations of motion of the individual particles is that the 
force $\bg(\bx)$ is bounded as $x \rightarrow 0$.

Introducing a top-hat window function $W(\bz=z_1, \dots, z_{d})$,  
\be
W(z_1, \dots, z_{d})=\left\{
        	  \begin{array}{ll}
		  1, &  \,{\rm if}\, |z_i| <  \frac{1}{2},\, \forall i=1\dots d \\
			0 & {\rm otherwise}
        	  \end{array}
	\right.
	\ee
we define the {\it coarse-grained} phase space density:
\be
f_0(\bx,\bv,t)\!= \!\!\!\int \!\!\frac{d^{d}\bx^\prime}{\lambda_x^d} \frac{d^d\bv^\prime}{\lambda_v^d} W\!\!\left(\!\frac{\bx-\bx'}{\lambda_x}\!\right)\!\! W\!\!\left(\!\frac{\bv-\bv'}{\lambda_v}\!\right)\!\! f_{k}(\bx'\!,\bv'\!,t)
\ee
where  $\lambda_x$ and $\lambda_v$ are the characteristic sizes of the coarse-graining cell 
in position and, respectively, velocity space. We will denote by $\mathcal C(\bx,\bv)$ the coarse 
graining cell centred at $(\bx,\bv)$, which thus has a phase space volume $\lambda_x^d \, \lambda_v^d$, and 
\be
N_{c}(\bx,\bv,t)=
\lambda_x^{d}\lambda_v^{d}f_0(\bx,\bv,t) 
\ee
is the number of particles in $\mathcal C(\bx,\bv)$.
We suppose that the coarse graining cell is always much smaller (in both real and
velocity space) than the characteristic size of the system, but sufficiently large to contain
a large number of particles, i.e.,  $\lambda_x \ll L_x$ and $\lambda_v \ll L_v$ where
$L_x$ and $L_v$ the characteristic size of the system in coordinate and
velocity space respectively, and 
\be
1 \ll N_{c}(\bx,\bv,t) \ll N \,.
\ee

We define also the coarse-grained 
spatial density as 
\be
n_0(\bx,t)=\int d^d \bv f_0(\bx,\bv,t)=\int \frac{d^{d}x^\prime}{\lambda_x}W \left(\frac{\bx-\bx'}{\lambda_x}\right)n_{k}(\bx')
\ee

By integrating Eq. (\ref{eq:Klimontovich}) over the coarse-graining cell we obtain straightforwardly
the following equation: 
\be \label{eq:finiteN_vlasov}
 \frac{\dd f_0 }{\dd t } + \bv \cdot \frac{\dd f_0 } {\dd \bx} + \bF_0 (\bx) \cdot \frac{\dd f_0 }{\dd \bv} \!=\! -\frac{\dd }{\dd \bx} [f_0\, 
 \xi_{\bv}]  -\frac{\dd }{\dd \bv} [f_0 \,\xi_{\bF}]
\ee
where 
 \be
 \label{mean-force}
 \bF_0(\bx,t)=\int_{\Omega}\bg(\bx-\bx')n_0(\bx',t)d^{d}\bx'\,,
 \ee  
i.e., the force at the point $\bx$ due to the coarse-grained distribution (which
we will identify as the mean-field force). Furthermore  
\be 
\label{eq:xiv}
\xi_v(\bx,\bv,t)=\left[\frac{1}{N_{c}(\bx,\bv,t)} \sum_{i \in \mathcal C(\bx,\bv)}\bv_{i}(t)\right]-\bv \, 
\ee
where $\bv_i$ is the velocity of particle $i$, and
\be \label{eq:xif}
\xi_{\bF}(\bx,\bv,t)= \left[\frac{1}{N_{c}(\bx,\bv,t)} \sum_{i\in\mathcal C(\bx,\bv)}\bF_i(t) \right] -\bF_0(\bx,t)
\ee
where $\bF_i$ is the exact force acting on the particle $i$, i.e., 
$\bF_i=\bF (\bx_i,t)$ where
\be
\bF (\bx,t)=\int_{\Omega}\bg(\bx-\bx')n_k(\bx',t)d^{d}\bx'\,.
\ee   
Thus $\xi_\bv$ is the ``velocity fluctuation" in the cell $ (\bx,\bv)$ around the coarse-grained velocity $\bv$, i.e., difference 
between the arithmetic mean of particles velocities in the cell (i.e. the velocity of the center of mass) and the velocity 
at the centre  of the coarse-graining cell, and  $\xi_{\bF}(\bx,\bv,t)$ is the ``force fluctuation" around the
coarse-grained force, i.e., the difference between the arithmetic mean of the exact forces acting
on each particle in the cell (equal to the force on the centre of mass of the particles in the cell)
and the force at the centre of the cell due to the coarse-grained particle distribution. 

If the right hand side of Eq. (\ref{eq:finiteN_vlasov}) is set equal to zero, we obtain,
given Eq. (\ref{mean-force}), the Vlasov equation for the coarse-grained phase space 
density $f_0 (\bx, \bv, t)$. Establishing the validity of the Vlasov equation in an
appropriate limit thus requires showing that the terms on the right-hand side may 
indeed be taken to zero in this limit. For a real system, for which $N$ is finite 
and the typical number of particles in a coarse-graining cell
is finite, Eq. (\ref{eq:finiteN_vlasov}) is not closed for the coarse-grained phase space
density, but rather coupled to the fine-grained density through the terms on
the right-hand side. If it is possible to define the Vlasov limit for the system,
these terms will represent at any finite (but large) $N$, small perturbations
to the pure Vlasov evolution of the coarse-grained distribution associated
with the ``graininess" of the system which, under suitable hypothesis, are 
responsible for the relaxation of the system to the thermodynamic equilibrium. 
In the rest of this article we focus
on these terms, and establish their scaling with $N$ (or, equivalently, as
a function of the characteristic scales of the coarse-graining cell),
given certain simplifying hypotheses. 

\section{Statistical evaluation of the finite-$N$ fluctuating terms}
\label{Statistical evaluation of the finite-$N$ fluctuating terms}

We now focus on the two terms on the right-hand side of Eq. (\ref{eq:finiteN_vlasov}).
Their direct evaluation is clearly impossible as in principle it requires knowledge of the full
fine-grained phase space density. We can, however, determine how they scale
as a function of relevant parameters by using a statistical approach:  given
a coarse-grained distribution $f_0(\bx, \bv)$ we can consider 
the fine-grained distribution to be characterized by an ensemble 
of realizations of particle distributions having $f_0(\bx, \bv)$ as 
mean density.
If we know the statistical properties of this
ensemble, we can then, in principle, calculate those of the fluctuating terms 
on the right-hand side of Eq.~\eqref{eq:finiteN_vlasov}.
In particular, we can then consider how the amplitudes of these
statistical quantities depend on the relevant parameters. 
 
We make here the most simple possible hypothesis about this
ensemble for the fine-grained phase space density: we suppose
that it corresponds to {\it the ensemble of realizations of an
inhomogeneous Poissonian point process with mean density
given by the coarse-grained phase space density} $f_0(\bx, \bv)$.
In other words we assume that the particles are randomly distributed, 
without any correlation, inside each coarse-grained cell, with a
mean density which varies from cell to cell following $f_0(\bx, \bv)$.
The density-density correlations are thus fully described by the 
one point distribution $f_0(\bx, \bv)$, and all other correlations, 
of the fluctuations around this mean density, are neglected. 
Physically this means we retain the ``pure" finite $N$ 
(discreteness) effects arising from the fluctuations of the 
mean density, but neglect the correlation of these fluctuations.
Alternatively one can consider that we proceed by assuming
complete ignorance of the distribution of particles
below the coarse-graining scale, other than the
information furnished about it by the coarse-grained
density itself. This allows us in effect to close, at
least in a statistical formulation,  Eq. (\ref{eq:finiteN_vlasov})
for $f_0(\bx, \bv)$. Indeed this hypothesis is
arguably the most natural one to make in seeking
to obtain a criterion for the validity of the 
Vlasov equation which is based only on the
 coarse graining density $f_0(\bx, \bv)$ and 
 the properties of the pair interaction itself.

Formally we can state our assumption to be that the
relevant terms can be evaluated by considering  
an ensemble of realizations of a point process with 
the N particle probability distribution
in phase space given by (see, e.g., 
\cite{gabrielli2006statistical})
\be \label{eq:NbodyPDF-unconstrained}
\mathcal{P}_{N} (\bx_1,\bv_1; ..;\bx_N, \bv_N) =\prod^{N}_{i=1} \frac{f_0 (\bx_i, \bv_i)}{N}
\ee
assuming the coarse-grained phase space density $f_0(\bx, \bv)$ 
to be a smooth function. In practice it is convenient to perform
the calculation with finite coarse-grained cells in which $f_0(\bx, \bv)$
is fixed, and the particles are distributed randomly in each
coarse-grained cell.

\subsection{Mean and variance of $\xi_\bv$}

We first evaluate the mean and variance of $\xi_v$, as defined by 
Eq.~(\ref{eq:xiv}), assuming now the defined properties of
the ensemble of realizations.  Given that the latter assigns 
randomly the velocities of the particles in the coarse-grained cell, which
is centred at $(\bx, \bv)$,  we evidently have
\be 
\label{eq:xiv-av}
\left< \frac{1}{N_{c}(\bx,\bv,t)} 
\sum_{i \in \mathcal C(\bx,\bv)}\bv_{i}(t) \right> =\bv \,, 
\ee
where $\left<...\right>$ denotes the ensemble average.
Therefore we have, as would be expected, 
\be
\left< \xi_{\bv} \right>=0\,.
\ee 

Calculating the variance of $\xi_{\bv}$ gives straightforwardly that
\be
\label{xi-v-variance}
\left<\xi_{\bv}^{2}\right> 
=\frac{1}{N_c(\bx,\bv)}\frac{d}{12} \lambda_v^{2}
\ee
where we can take $N_c=f_0(\bx, \bv)\lambda_x^d \lambda_v^d$ to be 
the average number of particles in the cell, given that  $N_c=\left<N_c\right>+\delta N_c$ 
where by hypothesis $\delta N_c\sim \sqrt {N_c}\ll N_c$
\footnote{In a uniform Poisson process with mean density $n_0$, 
the PDF $P(N,V)$ of the number of particles $N$ in a volume V is 
$P(N,V)= (n_o V)^N e^{-n_0 V}/N!$ (see, e.g., 
\cite{gabrielli2006statistical})}.
Given that $N_c (\bx,\bv)$ is large, and $\bv_i$ are considered independent
and identically distributed variables, $\xi_{\bv}$ is thus simply 
a Gaussian random variable with mean zero and 
variance given by Eq. (\ref{xi-v-variance}).

\subsection{Mean and variance of $\xi_\bF$}

Let us evaluate now the first two moments of $\xi_{\bF}$.
To evaluate these quantities in the ensemble defined above, it
is useful first to note that
\begin{align}
\label{force-sum}
\bar{\bF} (\bx,\bv, t)
&\doteq  \frac{1}{N_{c}(\bx,\bv,t)} \sum_{i\in\mathcal C(\bx,\bv)}\bF_i(t) \\  \nonumber
&= \frac{1}{N_{c}} \sum_{i=1}^{N_c} \sum_{I=1}^{N-N_c}  \bg (\bx_i-\by_I) 
\end{align}
where the $i$ runs over the $N_c$ particles {\it inside the coarse-graining cell}
$\mathcal C(\bx,\bv)$, and the index $I$ over the other $N-N_c$
particles {\it outside the cell}. This is the case because we assume
$\bg(-\bx)=-\bg(\bx)$ (making the sum over all the mutual forces of 
the pairs of particles in the cell $\mathcal C(\bx,\bv)$ vanishes).
As the individual pair forces depend only 
on the spatial positions of particles, we need only specify, to calculate 
the ensemble average of powers of the force, the 
probability distribution for the {\it spatial} 
positions of these points. 
Given the writing of the force in  (\ref{force-sum}), in which the sum is 
performed separately over the particles inside and outside the 
coarse-grained cell $\mathcal C(\bx,\bv)$ considered, it is 
convenient to write the ensemble average in a
similar form.  As noted above we can take
$N_c$ to be fixed and equal to its average value,
$n_0(\bx) \lambda_x^d$, up to negligible corrections 
of order $1/\sqrt{N_c}$. We can then write the N particle 
probability distribution as 
\be \label{eq:NbodyPDF}
\mathcal{P}_{N}^{\,\prime} (\bx_1,..\bx_{N_c}; \by_1,..\by_{N-N_c}) = \prod^{N_c}_{i=1} p(\bx_i) \prod^{N-N_c}_{I=1} \hat{p}(\by_I)
\ee
where $p(\bx)$ is the one-point probability distribution function of the
spatial position of a particle given the condition that it is contained in 
the cell, and $\hat{p} (\by)$ is the one-point probability distribution of the 
spatial position of a particle given the condition that it is outside 
the cell. Given that the $N_c$ particles are randomly distributed
inside the cell, we have simply
\be
p(\bx_i)=\left\{
        	  \begin{array}{ll}
		 \lambda_x^{-d} , & \bx \in \mS (\bx) \\
			0, & {\rm otherwise}
        	  \end{array}
	\right.
\ee
where $\mS (\bx)$ is the support of $W(\bx/\lambda_x)$, i.e.,
the ``stripe" in phase space with the same spatial coordinates as 
the phase space cell $\mathcal C(\bx,\bv)$ (see Fig.~1).
Assuming the number of particles in the cell $N_c$
to be small compared to the total number of particles $N_s(\bx, \bv)$ in 
this stripe (which is itself small compared to the total number
of particles $N$), $\hat{p} (\by)$ can be approximated
everywhere simply by the unconditional one point PDF for
the spatial distribution obtained by integrating 
(\ref{eq:NbodyPDF-unconstrained}) over all
but one space coordinate, i.e.,
\be
\hat{p}(\by)= \frac{n_0(\by)}{N}
\ee

\subsubsection{Mean of $\xi_{\bF}$}
Using (\ref{eq:NbodyPDF}) to calculate the ensemble average of the exact force exerted 
by all particles on those in a coarse-grained cell, we have
\bea
\left< \bar{\bF} (\bx,\bv, t) \right> 
&=& (N-N_c)  \int_\Omega d^d \bx^\prime \int_\Omega d^dy^\prime  p(\bx^\prime) \hat{p} (\by^\prime) \bg (\bx^\prime-\by^\prime)\nonumber \\
&=&   \int_{\mS (\bx)} \frac{d^d \bx^\prime}{\lambda_x^{d}} \int_\Omega d^d \by^\prime   n_0 (\by^\prime) \bg (\bx^\prime-\by^\prime)\nonumber \\
&=&   \int_{\mS (\bx)} \frac{d^d \bx^\prime}{\lambda_x^{d}} \bF_0 (\bx^\prime)
\eea
Thus
\be
\left< \xi_{\bF} (\bx,\bv, t) \right> 
=  \int_{\mS (\bx)} \frac{d^d \bx^\prime}{\lambda_x^{d}} [\bF_0 (\bx^\prime)- \bF_0 (\bx)]\,.
\ee
Assuming that we can neglect the variation of the mean-field $\bF_0$ in the 
coarse-graining cell, we obtain 
\be
\left< \xi_{\bF} (\bx,\bv, t) \right> 
= 0
\ee

\begin{figure} 
\includegraphics[scale=0.35]{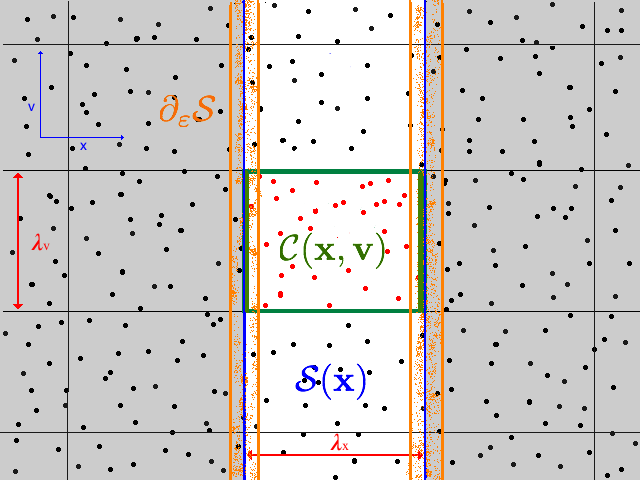}
\caption{(Color online) Schema, in one dimension, of the coarse-graining of phase space, showing
also  the different regions into which the domain of integration is divided for our analysis 
of the integrals in the force variance: $\mC$ is the coarse-grained cell with centre 
at $(\bx, \bv)$, $\mS$ (in white) is the ``stripe" enclosing all points in $[\bx-\frac{\lambda_x}{2},
\bx+\frac{\lambda_x}{2}]$; $\partial_\varepsilon \mS$ is the region containing particles within a 
distance $\varepsilon$ of the boundary of the stripe.}
\label{fig:schema}
\end{figure}

\subsubsection{Variance of $\xi_{\bF}$}
We now calculate 
\begin{align}
\bigl< \xi^2_F(\bx,\bv,t) \bigr> =
\left< \bar{\bF}^2(\bx,t)\right>-\bF_0^2(\bx,t)
\end{align}
We first break $\left< \xi^2_F(\bx,\bv,t) \right>$ into two parts
\be
\left< \bar{\bF}^2(\bx,t)\right>= \left< \frac{1}{N^2_{c}} \sum_{i} \bF^2 (\bx_i) \right>
+ \left< \frac{1}{N^2_{c}} \sum_{i, j; j \neq i} \bF (\bx_i) \cdot \bF (\bx_j) \right>
\ee
where all the sums are over the $N_c$ particles in the cell $\mathcal C (\bx, \bv)$.
We will refer to the first term on the right hand side  as the ``diagonal" contribution, and the second 
term as the ``off-diagonal" contribution to the variance: the first is the contribution to the variance
due to the variance of the force on each particle of the cell, the second the contribution to
the variance arising from the {\it correlation} of the forces on different particles in the cell.

Further we can use Eq. (\ref{force-sum}) to rewrite each term as two terms 
\bea 
\left< \bar{\bF}^2(\bx,t)\right> &=& \left< \frac{1}{N^2_{c}} \sum_{i} \sum_{I} \bg^2 (\bx_i-\bx_I) \right> \nonumber \\
&+& \left< \frac{1}{N^2_{c}} \sum_{i} \sum_{I,J; I\neq J} \bg (\bx_i-\bx_I) \cdot  \bg (\bx_i-\bx_J) \right> \nonumber \\
&+& \left< \frac{1}{N^2_{c}} \sum_{i,j; i\neq j}  \sum_{I} \bg (\bx_i-\bx_I) \cdot  \bg (\bx_j-\bx_I) \right> \nonumber \\
&+& \left< \frac{1}{N^2_{c}} \sum_{i, j; i \neq j}  \sum_{I,J; I\neq J} \bg (\bx_i-\bx_I) \cdot  \bg (\bx_j-\bx_J) \right> \nonumber
\eea
where $i,j$ again denote sums over the $N_c$ particles inside the coarse-grained cell
and $I,J$ over the $N-N_c$ particles outside the cell.
Performing the ensemble average by integrating over the PDF (\ref{eq:NbodyPDF}), the result
can be conveniently divided in two parts. The first and third terms give
\be 
\int_{\Omega} \!\! \frac{d^d\bx'}{\lambda_x^d} n_0(\bx') \int_{\mS(\bx)} d^d\by 
\,\bg^2(\by'-\bx')
\ee
and
\be
\left(1-\frac{1}{N_c(\bx,\bv)}\right) 
\int_{\Omega}\!\! \frac{d^d\bx'}{\lambda_x^d}
 n_0(\bx') \left[ \int_{\mS(\bx)} \!\!\!\!\!\!\!\! d^d\by' 
\bg(\by'-\bx') \right]^2 
\ee
respectively. Both the second and fourth terms can be expressed purely in terms of the mean-field,
as
\be 
\frac{1}{N_{c}} \left(1- \frac{1}{N}\right)  \int_{\mS (\bx)}\!\! \frac{d^d x^\prime}{\lambda_x^{d}} \bF_0 ^2(\bx^\prime)
\ee
and 
\be
\left(1- \frac{1}{N_{c}}\right)\!\! \left(1- \frac{1}{N}\right)\!\!  \int_{\mS (\bx)}\!\!\!\!\!\frac{d^d \bx_1}{\lambda_x^{d}} 
\int_{\mS (\bx)} \!\!\!\!\! \frac{d^d \bx_2}{\lambda_x^{d}} \bF_0(\bx_1) \cdot \bF_0(\bx_2)
\ee
respectively.

Assuming again that we can neglect the variation of the mean-field $\bF_0$ in the coarse-graining cell, 
we can perform the integrals in the last two expressions, and then obtain 
\begin{align} \label{eq:Var_Xi_F}
&\left< \xi^2_F(\bx,\bv) \right>= \frac{1}{N_c(\bx,\bv)} 
\int_{\Omega}\!\! d^d\bx' n_0(\bx') \int_{\mS(\bx,\lambda)} \!\!\!\!\!\!\!\!\!\!\!\!\!\! d^d\by 
\lambda_x^{-d} \bg^2(\by'-\bx')\nonumber \\
&+(1-\frac{1}{N_c(\bx,\bv)}) 
\int_{\Omega}\!\! d^d\bx'
 n_0(\bx') \left[ \int_{\mS(\bx,\lambda)} \!\!\!\!\!\!\!\!\!\!\!\!\!\! d^d\by' 
\lambda_x^{-d} \bg(\by'-\bx') \right]^2 \nonumber\\
&-\frac{1}{N} \bF^2_0(\bx)
\end{align}

Our analysis below will focus essentially on the first two terms in this expression
as they are those which describe the contribution to the fluctuating terms which
are potentially sensitive to the small scale properties of the pair force 
$\bg (\bx)$. We note that the first term comes from
the ``diagonal" part of the variance, and more specifically it represents
the contribution to the variance arising from the force on a single
particle in the cell due to a particle outside the cell: we will 
thus refer to it  as the {\it two body} contribution. The second integral in (\ref{eq:Var_Xi_F}),
on the other hand, arises from the ``off-diagonal" part of the variance,
and more specifically it represents the contribution to the variance
of the force on the cell due to the correlation of the force exerted on
two particles inside the cell exerted by a particle outside
the the cell: we will refer to it therefore as the {\it three body}
contribution.

\section{Parametric dependence of the fluctuations}
\label{Parametric dependence of the fluctuations}

We now analyse the expressions we have obtained, focussing on
how their value depends parametrically on the relevant 
parameters we have introduced, notably the number of
particles in the system ($N$), its size ($L_x$, $L_v$), the
number of particles in the coarse-graining cells ($N_c$)
and the coarse-graining scales ($\lambda_x$, $\lambda_v$).
Further we will take the pair force to be given by 
$\bg(\bx)=g|\bx |^{-\gamma}\hat{\bx}$ for $\gamma <0$, 
and by 
\be\label{pair-force-def}
\bg(\bx)=g\left\{
        	  \begin{array}{ll}
		|\bx |^{-\gamma}\hat{\bx} , & |\bx|\geq\varepsilon\\
			\varepsilon^{-\gamma}\;\hat{\bx}, & |\bx|<\varepsilon
        	  \end{array}
	\right.
	\ee
for $\gamma >0$, where $g$ is the coupling constant (and $g<0$ for the 
case of an attractive interaction). We will focus then on
how the amplitude of the fluctuations of the force depend
also on the exponent $\gamma$ and the characteristic length
$\varepsilon$ at which the force is regularized at small
scales. Indeed it is evident that we must introduce such 
a regularization of the pair force, as without it the integrals in 
(\ref{eq:Var_Xi_F}) can be ill-defined. It is important to
note that our essential scaling results are not dependent
on the use of the specific form of the regularization
in Eq. (\ref{pair-force-def}): what is necessary to obtain
these results is only that the force be bounded above by
a constant of order $\varepsilon^{-\gamma}$ below 
a scale of order $\varepsilon$.

\subsection{Mean field Vlasov limit} 

The mean-field Vlasov limit is formulated by taking
$N \rightarrow \infty$ at fixed system size, and scaling
$g \propto N^{-1}$ so that the mean field $\bF_0$ remains fixed.
Applying this procedure to the expressions we have obtained for 
$\left< \xi_\bv^2 \right>$ and $\left< \xi_\bv^2 \right>$, both indeed
converge to zero, for any non-zero $\varepsilon$.
In order to obtain these expressions we have, as noted,
assumed also that the mean-field does not vary on the scale of
the coarse-graining cell. Assuming that the characteristic scale of variation of the
coarse-grained phase space density, and mean field, 
is the system size $L_x$, for a {\it finite} coarse-graining cell
we expect corrections to our expressions due to the variation of 
these quantities on the scale $\lambda_x$ which are suppressed 
at least  by $\lambda_x/L_x$ relative to those we have calculated.
The coarse-grained phase space density $f_0 (\bx, \bv, t)$ 
thus indeed obeys the Vlasov equation in the usual
formulation of this limit, when the size of the 
coarse-graining cell is taken to be negligible with
respect to the system size.

We now study more closely the approach to the Vlasov limit 
as characterized by the scaling behaviour of the corrections to it.
We assume that these are given by those of the statistical quantities 
we have calculated, i.e., we take
\be 
|\xi_\bv| \sim \sqrt{\left< \xi_\bv^2 \right>}\,,\,\,\, |\xi_\bF| \sim \sqrt{\left< \xi_\bF^2 \right>}
\ee

\subsection{Velocity fluctuations}

Given the result (\ref{xi-v-variance}) and that $N_c \sim\frac{N}{L_x^d L_v^d}\lambda_x^d \lambda_v^d$
we infer
\be \label{eq:xiv_sim}
|\xi_\bv| \sim  \frac{1}{\sqrt{N}}\left(\frac{L_x}{\lambda_x}\right)^{d/2} \left(\frac{L_v}{\lambda_v}\right)^{d/2} \lambda_v
\ee
There is evidently no dependence on the pair force in this term. 
As already noted we can recover the Vlasov limit by taking $N \rightarrow \infty$.
Further we can see that this result remains valid for arbitrarily
small (but non-zero) values of the ratio $\frac{\lambda_x}{L_x}$ 
so that variation of coarse-grained
quantities on the coarse-grained scale can indeed be neglected.
 
\subsection{Force fluctuations}

The scaling of the last term on the right-hand side (\ref{eq:Var_Xi_F}) is already explicited, 
representing simply a fluctuation of the force about the mean field of order $1/\sqrt{N}$ times 
the mean field  itself.  In order to determine
 the dependences of the first two terms we need to analyse carefully that
of the integrals. To do so we divide the domains of integration in the double integral into
appropriate subdomains, isolating the region which may depends on the lower cut-off $\varepsilon$
in the pair force.

\subsubsection {2 body contribution}
We consider first the diagonal two body contribution, writing the double integral
as
\begin{eqnarray}
&&\left[ \int_{\Omega/S/\partial_{\varepsilon}\mS} d^d\bx'
+ \int_{\partial_{\varepsilon}\mS}   d^d\bx'
+\int_{S/\partial_{\varepsilon}\mS}d^d\bx'
\right]\otimes\nonumber\\
&&\left(n_0(\bx')
\int_{\mB(\bx', \varepsilon)}
d^d\by'\lambda_x^{-d} \bg^2(\by'-\bx')\right.\nonumber\\
&&\left.+ n_0(\bx')\!\!
\int_{\mS(\bx)/\mB(\bx',\vep)}
d^d\by \lambda_x^{-d} \bg^2(\by'-\bx') \right)
\label{integral-expanded}
\end{eqnarray}
where in this context $\otimes$ indicates the integration operation on the terms in square brackets.
As illustrated for the one dimensional case in Fig. \ref{fig:schema}, the integral over 
$\bx^\prime$, over the
whole of space ($\Omega$), has been divided into three parts:
\begin{itemize}
\item $\partial_{\varepsilon}\mS$: the set of points which are within a distance $\varepsilon$
of the boundary of the stripe $\mS$. The volume of this region is of order $\varepsilon \lambda_x^{d-1}$.
\item $\mS/\partial_{\varepsilon}\mS$: the set of points belonging to $\mS$ but not belonging to
$\partial_{\varepsilon}\mS$. For $\varepsilon \ll \lambda_x$, its volume is of order $\lambda_x^{d}$.
\item $\Omega/\mS/\partial_{\varepsilon}\mS$: the rest of space.
\end{itemize}
and the integral over $\by^\prime$, over the stripe $\mS$ has been divided into two parts:
\begin{itemize}
\item $B (\bx^\prime, {\varepsilon})$: the set of points in $\mS$ which are a distance of less than
$\varepsilon$ from the point $\bx^\prime$. This region has a volume of order $\varepsilon^d$.
\item $\mS/B (\bx^\prime, {\varepsilon})$: the rest of $\mS$; for $\varepsilon \ll \lambda_x$, its volume is of order 
$\lambda_x^{d}$.
\end{itemize}
 
We consider now one by one the terms in the integral written as in (\ref{integral-expanded}).
The first term in the integration over $\bx^\prime$ excludes the region where the
pair force is $\varepsilon$-dependent, and is over a volume of order $L_x^d$. Hence if 
we suppose $n_0\sim const$, it gives a contribution which scales as 
\be
\sim n_0 L_x^{d-2\gamma}
\ee

For the second region of integration over $\bx^\prime$, of volume of order $\varepsilon\lambda_x^{d-1}$,
the region $\mB(\bx',\vep)$ of the integration over $\by^\prime$ gives a contribution of
order $\varepsilon^{-2\gamma}$ over a volume of order $\varepsilon^d$, and thus
\be
\sim n_0 (\varepsilon\lambda_x^{d-1}) \varepsilon^d \lambda_x^{-d} \varepsilon^{-2\gamma}\,.
\ee
In the region $\mS(\bx)/\mB(\bx',\vep)$ on the other hand, of volume of order $\lambda_x^d$,
we have 
\be
\sim n_0 (\varepsilon\lambda_x^{d-1}) \lambda_x^{-2\gamma}
\ee

In the third region of integration over $\bx^\prime$, of volume of order $\lambda_x^d$, 
we obtain again a contribution of order $\varepsilon^{-2\gamma}$ in the volume
$\mB(\bx',\vep)$, and thus
\be
\sim n_0 \vep^{d-2\gamma}
\ee
while the second term region of the $\by^\prime$ integration gives
\be
\sim n_0 \lambda_x^{d-2\gamma}\,.
\ee

\subsubsection{3 body contribution}

Proceeding in the same manner we write the 3-body contribution to the variance
of the force as
\begin{eqnarray}
&&\left[ \int_{\Omega/S/\partial_{\varepsilon}\mS}d^d\bx'
+ \int_{\partial_{\varepsilon}\mS}d^d\bx'
+\int_{\mS/\partial_{\vep}\mS}d^d\bx'
\right]\otimes\nonumber\\
&& \left(n_0(\bx')\int_{\mB(\bx', \varepsilon)}
d^d\by'\lambda_x^{-d} \bg(\by'-\bx') \right. \nonumber\\
&&\left.+n_0(\bx')\int_{\mS(\bx,\lambda)/\mB(\bx',\vep)}
d^d\by \lambda_x^{-d} \bg(\by'-\bx') \right)^2
\nonumber
\end{eqnarray}

The first term in the integration over $\bx^\prime$ gives then a contribution 
\[
\sim n_0 L_x^{d-2\gamma} 
\] 
Compared to the 2-body integral, the analysis of the remaining parts 
is essentially the same, except for one important difference: as the
integration over $\by^\prime$ is over the vector pair force, the 
integral over $\by^\prime$ is zero when integrated in a sphere
around $\bx^\prime$; in particular the integration over $\mB(\bx',\vep)$ 
vanishes when $\mB(\bx',\vep)$ is fully contained in $\mS$. This 
is the case for the integration region $\mS/\partial_{\vep}\mS$,
which therefore does not depend on $\varepsilon$ and simply
gives a contribution of order
\be
\sim n_0 \lambda_x^{d-2\gamma}\,.
\ee 

For the second integration region in the integral over $\bx^\prime$, of volume 
$\sim \varepsilon \lambda_x^{d-1}$, 
$\mB(\bx',\vep)$ is not fully contained in $\mS$ and we have therefore
a contribution from this part of the integration over $\by^\prime$ of
order $\varepsilon^d \lambda_x^{-d} \varepsilon^{-\gamma}$, while
the second part, which does not depend on $\varepsilon$, gives a 
contribution of order $\lambda_x^{-\gamma}$.
Taking the square and multiplying by the volume of $\partial_{\vep}\mS$,
we obtain three terms: 
\be
\sim n_0 \varepsilon \lambda_x^{d-1} \varepsilon^{2d} \lambda_x^{-2d} \varepsilon^{-2\gamma}
\ee 
from the square of the first term, 
\be\label{cross}
\sim n_0 \varepsilon \lambda_x^{d-1} \varepsilon^{d} \lambda_x^{-d} \varepsilon^{-\gamma}\lambda_x^{-\gamma}
\ee 
from the cross term, and 
\be
\sim n_0 \varepsilon \lambda_x^{d-1} \lambda_x^{-2\gamma}
\ee 
from the square of the second term. 

\section{Force fluctuations about the Vlasov limit:  dependence on $\varepsilon$}
\label{Force fluctuations about the Vlasov limit}

Gathering together the expressions derived above, we obtain, 
keeping only the leading divergence in $\varepsilon$ in each
of the 2-body and 3-body contributions,
\begin{align}  \label{eq:leading}
&\left< \xi^2_F(\bx,\bv) \right>\!\!=\!\! \frac{g^2}{N_c} 
\!\!\left(\! C_{\Omega}\, n_0 L_x^{d-2\gamma}
\!+\!C_{\mS} \,n_0 \lambda_x^{d-2\gamma}
 \!+\!C_{\partial \mS}\, n_0 \vep^{d-2\gamma} 
\right)
\nonumber \\
&+g^2(1-\frac{1}{N_c}) 
\left( 
C'_{\Omega}\, n_0 L_x^{d-2\gamma} 
+C'_{\mS}\, n_0 \lambda_x^{d-2\gamma}\right.  
\nonumber\\
&\left.
+C'_{\partial \mS}\, n_0 \varepsilon^{2d+1-2\gamma} \lambda_x^{-(d+1)} 
\right)
-\frac{1}{N} |\bF[n_0](\bx)|^2 \nonumber\\
\end{align}
where all $C_{*}$and $C'_{*}$ are constants (Note that we have
not included (\ref{cross}) because when it diverges, for
$\gamma > d+1$, the term retained is indeed more
rapidly divergent.)

Depending on the values of $\gamma$ and $d$ different terms 
dominate. We consider each case.

\subsection{Case $ \gamma<  \frac{d}{2}$}
In this range there are no divergences as $\varepsilon \rightarrow 0$, 
and for  $\varepsilon \ll \lambda_x \ll L_x$, the dominant term from
the two integrals is
\be 
g^2 n_0 L_x^{d-2\ga} \sim \frac{1}{n_0L_x}g^2n^2_0L_x^2\sim\frac{1}{N}|\bF_0 |^2 
\ee 
and therefore we infer the scaling of the total force fluctuation is 
\be
|\xi_{\bF}| \sim \frac{1}{\sqrt{N}} |\bF_0|
\ee
As noted above we therefore obtain in this case the Vlasov limit taking
$N \rightarrow \infty$ with $ g \sim 1/N$. Further we conclude that,
at finite $N$, the fluctuations around the mean-field force are 
dominated by contributions coming from fluctuations of the
density at the scale of the system size, which dominate
those coming both from the scale $\lambda_x$ of the coarse-graining
cell and those from the scale $\varepsilon$ at which the pair force
is regularized.

\subsection{Case $ \frac{d}{2} < \gamma<  d+\frac{1}{2}$}

In this range of $\gamma$, there is a divergence at $\vep \rightarrow 0$ in the
contribution coming from the 2-body term, while the 3-body term remains finite.
Keeping only the dominant contributions to the two integrals 
when $\varepsilon \ll \lambda_x \ll L_x$, we obtain
\begin{equation} \label{eq:dominant2}
\langle\xi_{\bF}^2 (\bx) \rangle
\sim 
C_{\vep} \frac{g^2 n_0(\bx)}{N_c(\bx,\bv)} \vep^{d-2\ga}
+ C'_{\lambda} g^2 n_0(\bx) \lambda_x^{d-2\ga} 
\end{equation}
where $C_{\vep}$ and $C'_{\lambda}$ are constants. 

Given the divergence in $\varepsilon$ we see explicitly that 
in this case the Vlasov limit is obtained taking $N \rightarrow \infty$ with
$g \sim N^{-1}$ at finite non-zero $\varepsilon$, and this limit
can only be defined if such a small scale regularisation
of the pair force is introduced.

Using $N_c(\bx,\bv)\propto f_0(\bx,\bv)\lambda_x^d \lambda_v^d$, we can
write the dominant fluctuations as
\be \label{eq:distiction}
\langle \xi_{\bF}^2(\bx)\rangle 
\sim g^2  \frac{n_0(\bx)}{\lambda_x^d} \left(C_{\vep}\frac{\vep^{d-2\ga}}{f_0(\bx,\bv)\lambda_v^d}+ C'_{\lambda} \lambda_x^{2d-2\ga} \right)
\ee 

This expression allows us to conclude, as anticipated, that there is a crucial difference 
between the following sub-cases: i) the range of $\gamma$ in which the first 
term dominates, and ii) the range in which the second term dominates:

\subsubsection{For $\frac{d}{2}<\gamma<d$ :}

In this case the exponent of $\lambda_x$ in the second term inside the brackets 
in (\ref{eq:distiction}) is positive, and therefore when we increase  $\lambda_x$
at fixed $\varepsilon$ (and fixed $\lambda_v$), this term dominates over 
the first one. More specifically when $\lambda_x \gg \vep (f_0(\bx,\bv)\lambda_v^d\vep^d)^{-\frac{1}{2d-2\gamma}}$ this term dominates, and 
\be
\langle\xi^2_F\rangle\sim g^2 n_0(\bx)\lambda_x^{d-2\gamma}
\ee 
Thus, even though the amplitude of the fluctuations depend on $\varepsilon$,
and diverges as $\varepsilon \rightarrow 0$, for a sufficiently large coarse-graining
cell the fluctuations become in practice effectively insensitive to the value
of $\varepsilon$,  for a wide range of values which is such that the larger 
is $\lambda_x$ the smaller is the lower limit on $\varepsilon$, with
the latter vanishing as $\lambda_x\to \infty$. Note that since
\begin{eqnarray}
&&|\xi_{\bF}|
 \sim \frac{1}{\sqrt{n_0(\bx)\lambda_x^d}}\left(\frac{\lambda_x}{L_x}\right)^{d-\gamma}|\bF_0(\bx)|
 \nonumber\\
 &&\sim \frac{1}{\sqrt{N}}\left(\frac{\lambda_x}{L_x}\right)^{d/2-\gamma}|\bF_0|\nonumber
\end{eqnarray}
we can also neglect the final term in (\ref{eq:Var_Xi_F}).

\subsubsection{For $d<\gamma<d+\frac{1}{2}$ :} 
In this range it is instead the exponent of $\lambda_x$ in the second term 
inside the brackets in (\ref{eq:distiction}) which is negative, and as a 
consequence it is now the first term which dominates when we
make the coarse-graining scale $\lambda_x$ large.  We
have therefore
\be
\langle \xi^2_\bF \rangle \sim g^2\frac{n_0}{f_0\lambda_x^d\lambda_v^d}\vep^{d-2\ga}\,.
\ee
Further as this can be rewritten as
\be
|\xi_{\bF}|
\sim \frac{1}{\sqrt{N N_c}} \left(\frac{\vep}{L_x}\right)^ {d/2-\gamma} |\bF_0|
\ee
it follows that for sufficiently small $\varepsilon$ this is the dominant contribution
to the fluctuations. In this range therefore the leading contribution to the force fluctuations is 
directly dependent on $\varepsilon$.

\subsection{Case $d+\frac{1}{2}<\gamma$}

In this case both the integrals giving the 2-body and 3-body contributions are
divergent at small $\varepsilon$, but the dominant divergence comes from the
latter giving
\be
\langle \xi^2_F (\bx)\rangle 
\sim
g^2 n_0(\bx)\frac{\vep^{2d+1-2\ga} }{\lambda_x^{d+1}} 
\ee
This case is therefore like the previous case ($d < \gamma < d+\frac{1}{2}$):
the dominant contribution to the fluctuations is divergent as $\vep \rightarrow 0$.

\section{Exact one dimensional calculation and numerical results}
\label{Exact one dimensional calculation and numerical results}

In the case of a one dimensional system, $d=1$, it is possible to
perform explicitly the integrals in (\ref{eq:Var_Xi_F}) to obtain exactly
the expression of the force fluctuations. In order to illustrate our main result 
above, we can compare the expressions we obtain, and in particular 
their leading scaling behaviours, with what is obtained directly by measuring the
force fluctuations in cells on realizations of a homogeneous Poisson particle 
distribution. As we are interested primarily in the $\varepsilon$ dependence
of these fluctuations we consider solely the contribution
to them from the part of the integral which is potentially
sensitive to them. We can therefore write 
\begin{align} \label{eq:Var_Xi_F_1d}
&\langle \xi_{\bF}^2 (\bx) \rangle_{\mS} =
\frac{n_0}{N_c} 
\int_{-\frac{\lambda_x}{2}}^{\frac{\lambda_x}{2}} dx'  \int_{-\frac{\lambda_x}{2}}^{\frac{\lambda_x}{2}} dy' 
\lambda_x^{-1} g^2(y'-x')\nonumber \\
&+n_0 (1-\frac{1}{N_c}) 
\int_{-\frac{\lambda_x}{2}}^{\frac{\lambda_x}{2}} dx'
\left[ \int_{-\frac{\lambda_x}{2}}^{\frac{\lambda_x}{2}} dy' 
\lambda_x^{-1} g(y'-x') \right]^2 \nonumber
\end{align}
where we use the subscript  in $\langle \xi_{\bF}^2 (\bx) \rangle_{\mS}$
to indicate that this is the contribution to the force variance sourced
by particles in the phase space ``stripe'' $\mS$, and
\be\label{pair-force-def-1d}
g(x)=g\left\{
        	  \begin{array}{ll}
		  \frac{x}{ |x |^{\gamma+1}}, & |x|\geq\varepsilon\\
			\frac{x}{ |x |} \varepsilon^{-\gamma}\, & |x|<\varepsilon
        	  \end{array}
	\right.
\ee
Integrating we obtain 
\begin{align} \label{exact-1d-force-in-stripe}
&\langle \xi_{\bF}^2 (\bx) \rangle_{\mS} =
\\ \nonumber
&g^2\frac{1}{N_c} \frac{n_0(\bx) }{\lambda_x} 
\left[
\frac{1}{1-\gamma}\varepsilon^{2-2\gamma}
-\frac{4}{1-2\gamma}\lambda_x\varepsilon^{1-2\gamma} 
\right.
\\ \nonumber
&+
\left.
\frac{1}{(1-2\gamma)(1-\gamma)}\lambda_x^{2-2\gamma}
\right]
\\ \nonumber
&+g^2(1-\frac{1}{N_c}) \frac{n_0(\bx)}{\lambda_x^2} 
\left[
\frac{2\gamma(1+2\gamma)}{2(1-\gamma)(3-2\gamma)}\varepsilon^{3-2\gamma}
\right.
\\ \nonumber
&-\frac{4\gamma}{(1-\gamma)^2(2-\gamma)}\varepsilon^{1-\gamma}(\lambda_x^{2-\gamma}-(\lambda_x-\varepsilon)^{2-\gamma})
\\ \nonumber
&-
\frac{4}{(1-\gamma)(2-\gamma)(3-\gamma)}\varepsilon^{-\gamma}(\lambda_x^{3-\gamma}-(\lambda_x-\varepsilon)^{3-\gamma})
\\ \nonumber
&+\frac{4}{(1-\gamma)(2-\gamma)}\varepsilon^{1-\gamma}(\lambda_x-\varepsilon)^{2-\gamma}
\\ \nonumber
&-\frac{4}{(1-\gamma)^2}\left(\frac{\lambda_x}{2} \right)^{2-2\gamma}
\!\!\!\left(\frac{\lambda_x}{2}-\varepsilon\right)
\!\!\!\sum^{\infty}_{n=0}\frac{(\gamma-1)_n}{1+2n}\frac{1}{n!}
\!\!\left(1-2\frac{\varepsilon}{\lambda_x}\right)^{2n}
\\ \nonumber
&+
\left.
\frac{2}{(1-\gamma)^2(3-2\gamma)}\lambda_x^{3-2\gamma}
\right]\\ \nonumber
\end{align}
where $(x)_n$ is the Pochhammer symbol,
$(x)_n =\frac{\Gamma (x+n)}{\Gamma(x)}=x(x+1)...(x+n-1)$.

Expanding this expression in the limit $\varepsilon/\lambda\rightarrow 0$,
and keeping only the leading $\varepsilon$-dependent terms, we obtain:
\begin{align} \label{variance-1d-expanded}
&\langle \xi_{\bF}^2 (\bx) \rangle_{\mS} = 
\\ \nonumber
&g^2 \frac{n_0(\bx)}{N_c} 
\left[
-\frac{4}{1-2\gamma}\varepsilon^{1-2\gamma}
+
\frac{1}{(1-2\gamma)(1-\gamma)}\lambda_x^{1-2\gamma}
\right]
\\ \nonumber
&+g^2 n_0(\bx) (1-\frac{1}{N_c}) 
\left[
\frac{2\gamma(1+2\gamma)}{2(1-\gamma)(3-2\gamma)}\lambda_x^{-2}\varepsilon^{3-2\gamma}
\right.
\\ \nonumber
&\left. +\frac{4}{(1-\ga)^2}\lambda_x^{-1-\ga}\varepsilon^{2-\ga}
+C \lambda_x^{1-2\gamma} \right]\\ \nonumber
\end{align}
where $C$ is a constant depending on $\gamma$. Comparing
this expression with (\ref{eq:leading}) which we obtained for
the $d$-dimensional case, we note that we indeed have 
agreement when we take $d=1$. The term of order $\varepsilon^{2-\ga}$
corresponds to the term of order $\varepsilon^{d+1-\ga}$ in (\ref{cross})
which we did not include in (\ref{eq:leading}) because it is never
the leading divergence.

We now compare these analytical expressions with those obtained 
from a direct numerical estimation of the same quantity
in a Poissonian realization of a particle system. To do so
we have distributed $N=72900$ particles randomly in a 
``phase space box"  (cf. illustration in Fig. \ref{fig:schema}) 
of  side $27 \lambda_v$ (in velocity) and $9 \lambda_x$ (in
position). Thus there are $243$ cells, containing, on average,
$300$ particles. 
We then calculate the exact force on each particle in
the system due to all particles in the stripe to
which it belongs, i.e., $\bF(x_{i})=\sum_{j\in \mathcal S}\bg(x_{i}-x_{j})$. 
For each cell we then average this quantity to get 
$\bar \bF=\frac{1}{N_{c}}\sum_{i \in \mathcal C}\bF_{k}(x_{i})$,
and we finally estimate the variance of this quantity over
the $243$ cells. 

\begin{figure}
\includegraphics[scale=0.22]{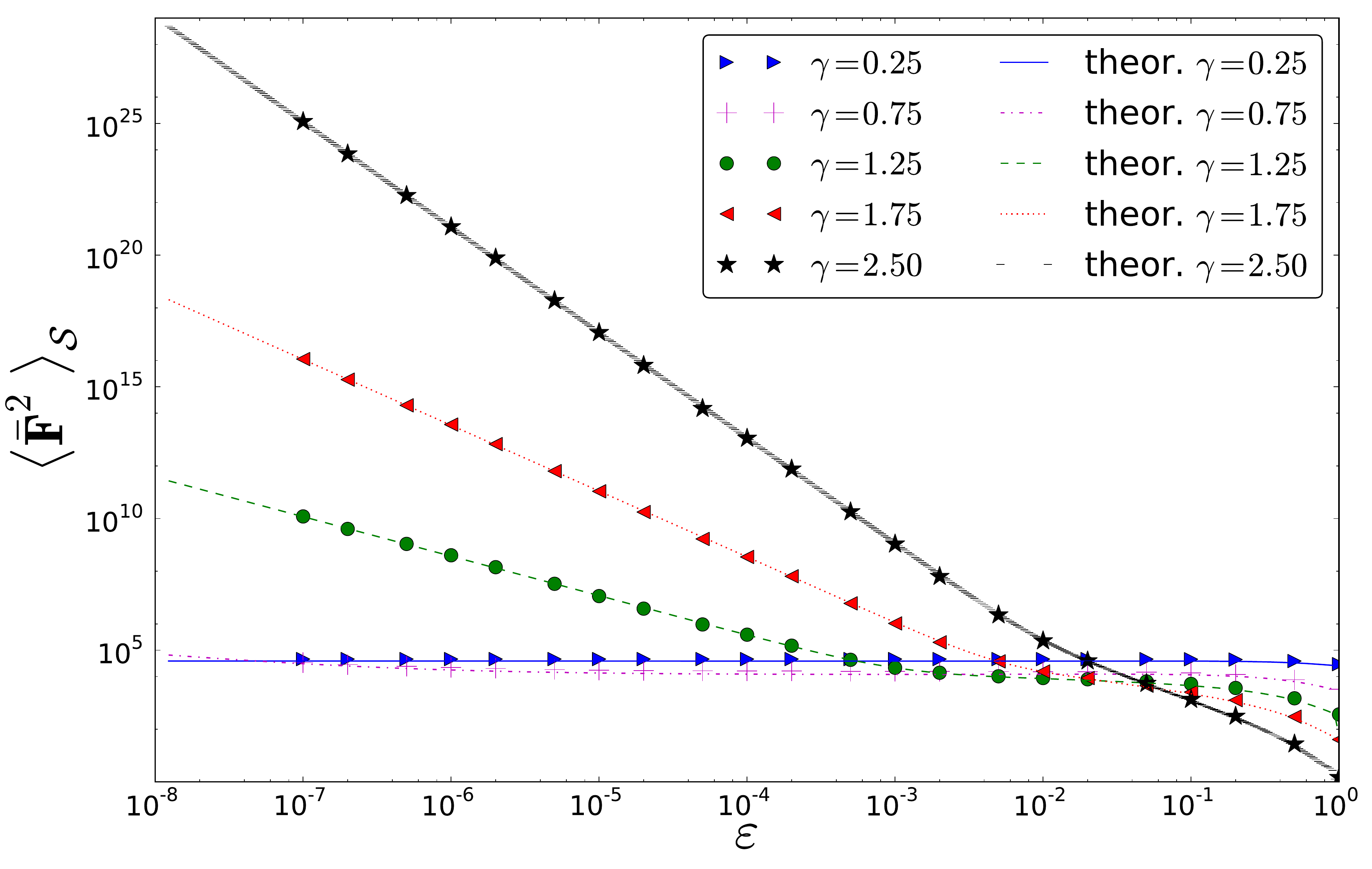}
\caption{(Color online) The variance of the force on particles in a phase space cell due to
particles in the corresponding stripe, plotted as a function of $\varepsilon$, for the 
different values of $\gamma$. The crosses indicate the points obtained from the 
numerical simulation described in the text and dotted lines are the 
corresponding analytical results, Eq. (\ref{exact-1d-force-in-stripe}). We use
units in which $\lambda_x=1$ and $g=1$.}
\label{fig:theovssim}
\end{figure}

Shown in Fig.~\ref{fig:theovssim} are the results of  these numerical simulations
compared with the theoretical results, Eq. (\ref{exact-1d-force-in-stripe}), for 
a range of values of $\varepsilon$ and different values of $\ga$. The 
agreement is excellent in all cases. Further we have verified that, 
except in the region where $\varepsilon$ approaches $\lambda_x$, the
results are in excellent agreement with the expression  (\ref{variance-1d-expanded}) 
for the  small $\varepsilon/\lambda_x$ behaviour.
Thus we find results completely in line with our determination
of the scalings in this limit for the $d$-dimensional case, leading
to the different behaviours described in 
Section \ref{Force fluctuations about the Vlasov limit}:
for $\gamma=0.25$, in the range $\gamma < d/2$, the force fluctuations
are independent of $\varepsilon$; for $\gamma=0.75$ and $\gamma=1.25$,
in the range $d/2 < \gamma < d/2 +1$, the divergent 2-body term $\sim
\varepsilon^{1-2\ga}$ dominates at small $\varepsilon$,  but is then
overtaken at larger $\varepsilon$ by the flat behaviour of the the 
dominant $\varepsilon$-independent term in the 3-body term;
for $\gamma=1.75$ and $\gamma=2.5$, both in the range 
$d/2 +1 < \gamma$ we have again at the smallest $\epsilon$
the behaviour  $\sim \varepsilon^{1-2\ga}$ from the 2-body
term, but at larger $\epsilon$ instead the dominant behaviour
$\varepsilon$-dependent term $\sim \varepsilon^{3-2\ga}$ from 
the 3-body term. 

\section{Discussion and conclusions}

Let us now discuss further the physical significance of these
results, and in particular how they justify the basic qualitative
distinction between interactions as we have anticipated in
the introduction. We have considered a generic 
$N$-particle system with Hamiltonian dynamics described by a 
two body interaction with a pair force $\sim 1/r^\ga$
and regularized below a scale $\varepsilon$ to a
constant value. Introducing a coarse-graining in phase 
space, one obtains an equation for the coarse-grained
phase space density with terms corresponding to
the Vlasov equation,  in addition to
``non-Vlasov'' terms which are functionals of
the microscopic phase density. We then made the 
hypothesis that the typical amplitude of these terms 
can be estimated by assuming this microscopic phase 
space density to be given by a realization of an inhomogeneous 
Poissonian point process, in which the mean
density is specified by the coarse-grained phase 
space density but the density fluctuations are uncorrelated.
In other words we neglect all contributions due to correlation 
of density fluctuations with respect to the ones due to the 
simple products of mean densities. 
Doing so 
we have determined the scalings of these latter terms
as a function of the different scales in the problem,
of $\gamma$ and of the spatial dimension $d$.
If the assumed hypothesis is valid (for large $N$), a limit 
in which the Vlasov equation applies is simply any one in 
which the derived non-Vlasov 
terms go to zero while the Vlasov terms tend to fixed finite values.
Our results show firstly that, for {\it any} interaction
in the class we have considered, such a limit
exists and can be defined in different ways:
notably, by taking $N$ to infinity at fixed values  
of the other scales, and in particular at fixed 
smoothing; or, alternatively, by taking the  
limit in which the coarse-graining scale, 
while always remaining small compared to 
the system size, is taken arbitrarily large
compared to the other scales, and in 
particular the smoothing scale $\varepsilon$. 
However, when this limit is taken, there is a
important difference between the case in
which $\gamma > d$ or $\gamma < d$:
for the former case the dominant correction
term to the Vlasov limit is always strongly 
dependent on $\varepsilon$, while in the latter case
the dominant term, at fixed $\lambda_x$, 
is independent of $\varepsilon$ in a wide range 
whose lower limit vanishes as $\lambda_x$ 
diverges. 
This different leading 
dependence on $\varepsilon$ corresponds
to fluctuations which are sourced by quite
different contributions in the two cases:
for $\gamma > d$, the dominant fluctuation
in the force on a coarse-graining cell comes
from the contributions coming from forces
on individual particles due to single particles
which are ``closeby'', i.e., within a radius
$\varepsilon$: for $\gamma < d$, on the
other hand, the dominant fluctuation
comes from the coherent effect on
two different particles in the cell 
coming from a single particle which
is ``far-away'', i.e. at a distance
of order the size of the coarse-graining
cell, or, even for $\gamma < d/2$, of order
the size of the system. Given that these 
amplitudes are those of the terms describing
the corrections to the Vlasov evolution
due to the large, but finite,  number of 
particles in a real system, this means 
that the dynamics of the particle system
at a coarse-grained scale in the two
cases is dominated by completely
different contributions: by the
particle distribution at the smallest
scales when $\ga > d$, and by
the particle distribution at the 
coarse-graining scale or larger
when $\ga < d$. Thus in the latter
case we can decouple the dynamics 
at the coarse-graining scale from that at 
smaller scales (the interparticle distance, 
the scale of particle size), while
for $\ga > d$ we cannot do so.
Or, in the language of 
renormalisation theory, the former admit
a kind of universality in which the
coarse-grained dynamics is insensitive
to the form of the interaction below
this scale, while the latter do not.
This is a basic qualitative
distinction between the dynamics in these
two cases, which corresponds naturally
to what one call ``short-range'' or
``long-range". In order to distinguish
it from the canonical distinction based
on thermodynamical considerations,
following \citep{Gabrielli2010}, we can
refer to $\gamma >d$ as {\it dynamically
 short-range}, and $\ga < d$ as 
 {\it dynamically long-range} 
 \footnote{Or, alternatively, if one adopts
 the terminology advocated by 
  \cite{Bouchet2010}, in which
the thermodynamic distinction is made
between ``strongly long-range" and
``weakly long-range", our classification
could be described as a distinction
between ``dynamically strongly long-range"
and ``dynamically weakly long-range" interactions.}.

For what concerns quasi-stationary states the implications
of this result and relevance of the classification are simple:
for all cases one would expect that such states may
exist (since the Vlasov limit exists) but the conditions
for their existence, which requires that the time scales
of their persistence be long compared to the system
dynamical time, will be very different.  For $\gamma > d$, 
their lifetime, which would be expected to be
directly related to the amplitude of the non-Vlasov
term, can be long only if the smoothing scale in
the force is sufficiently large;  in the 
case $\gamma < d$, their lifetime will be expected
to be independent of $\varepsilon$. We note that
this is precisely in line with results of
analytical calculations based on the 
Chandrasekhar approach to estimation of
the relaxation rate, and the results of numerical
simulations of systems of this kind reported
in \citep{Gabrielli2010a}.

These results are all, as we have emphasized, built on our
central hypothesis that correlation of density
fluctuations, associated with the finite particle number,
may be neglected. To formulate it 
we must define a smooth mean phase space
density by introducing a coarse-graining scale,
which is assumed to be a ``mesoscopic scale":
arbitrarily small compared to the system size, and 
yet large enough so that the phase space cell contains
many particles. Our central hypothesis is not
one of which we have proven the validity, but it is a 
consistent, simple and physically reasonable one,  analogous to
that of ``molecular chaos'' in the derivation of the 
Boltzmann equation. It is also in line 
with the fact that when the Vlasov approximation is valid, the 
system dynamics is defined in terms only of the mean density 
and fluctuations due to higher order density correlations are 
considered subdominant.
We note above all that it leads to conclusions and
behaviours which are very reasonable physically,
even in the cases we have not focussed on but
to which our analysis can be applied.
For example, if we consider a hard
repulsive core interaction without a smoothing,
i.e. let $\varepsilon \rightarrow 0$ for $\gamma$ 
large, we infer that the force fluctuation on
a coarse-grained cell diverges and that there 
is thus no Vlasov limit. Indeed in this case
an appropriate two body collision operator
is required
to take into account
the effects of interactions between particles.

Finally we note that 
the ensemble we have assumed to describe the 
fine-grained phase space density is a realization
of a Poisson process with {\it mean} density given by the 
coarse-grained space. This implies that we include
(Poissonian) fluctuations of particle number at all scales:
indeed these (finite particle number) fluctuations are 
the source of the fluctuation terms in the dynamical
equations which we have analysed. One could consider 
that it might be more appropriate physically to take, in
estimating the fluctuations induced at a given
coarse-grained density, the ensemble
in which the particle number is constrained
to be fixed in {\it all} coarse-graining cells. 
In other words we could average, given
a coarse-grained phase space density,
only over the configurations in which the
particles, of fixed number in each cell,
are distributed randomly within the
cells. It is straightforward to verify, with the
appropriate modification of the average,  
that doing so can only change our results 
for what concerns the large scale contributions 
to the force fluctuations: the
diverging behaviours at small scales we have
focussed on arise from the fact that there is
a finite probability for a particle to be
arbitrarily close to another one, and the
local value of the density will at most modify
the amplitude but not the parametric dependence
of this term.
On the other hand, our determinations
of the parametric dependence of the contributions
to the force fluctuations from the bulk will be 
expected to depend on how the particle fluctuations
at larger scales are constrained
\footnote{A simple
example is the case of one dimensional gravity, i.e., 
$\gamma=0$ in $d=1$. In this case Poissonian 
force fluctuations indeed diverge with system size 
\cite{Gabrielli2010b} precisely as derived here. 
On the other hand, because the
force is independent of distance, the force fluctuation
in a cell coming from particles in other cells vanishes
if the number of particles in these cells does not fluctuate.}.

In future work we plan to explore the possibility of developing
the approach used in this article to understand and describe further
the effect of the ``non-Vlasov" --- collisional --- terms on
the evolution of a finite $N$ system, i.e., to use it
to develop a kinetic theory for the $N$ particle
system. In this context it would be interesting to
determine whether existing kinetic equations such as 
that of Lenard-Balescu, or variants of it developed in 
the literature (see, e.g., \citep{Chavanis2012a, Chavanis2013} for
a discussion and references), could be derived 
in a different way from this starting point, or
even potentially modified or formulated differently.
In this respect we note that the results derived
here already provide a better basis for many derivations
of such equations (and in particular the Lenard-Balescu
equation) which take as a starting point the assumption
that the Vlasov equation applies in the large $N$
limit (and then derive the kinetic equation for perturbations 
about it). It would be interesting also to clarify in 
particular the relation between this approach and that of 
Chandrasekhar which, as we have noted, when 
extended to a generic interaction has been shown
\cite{Gabrielli2010a} to give very consistent conclusions 
about the sensitivity of collisional relaxation of 
quasi-stationary states to the small scale regularisation.   
Likewise it would be interesting to try to test directly
in numerical simulations for the validity of our central
hypothesis about correlations, and also characterize
using analytical methods the robustness of our central
conclusions to the existence of different weak correlations
of the density fluctuations.   

We thank Bruno Marcos and Pascal Viot for very useful
discussions.

%

\end{document}